# Synthesis of fine particles of a geometrically frustrated spin-chain system $Ca_3Co_2O_6$ through a pyrophoric route and its magnetic behavior

K. Mukherjee and E.V. Sampathkumaran

*Tata Institute of Fundamental Research, Homi Bhabha Road, Colaba, Mumbai 400005, India*

We report the synthesis of fine particles of a well-known geometrically frustrated spin-chain compound $Ca_3Co_2O_6$ through a new route, namely, a pyrophoric method employing triethanolamine (TEA) and studied the magnetic behavior of such specimens. We find that this method of synthesis yields particles made up of rods whose diameter appears to be controllable by the fraction of TEA during synthesis. It is seen that the two well-known magnetic transitions (~24 K and ~8 K) remains unaffected for all TEA concentrations used. The most notable finding is that the multi-step feature in the *M* versus magnetic-field isotherm reported for the single crystalline form tends to smoothen out gradually with decreasing rod thickness (from about 1 μm to a few hundred nm). Further, unlike for the single crystalline form known in the literature, there is no tendency for the relaxation time ($\tau$) to remain constant at low temperatures (<10 K) and $\tau$ remains temperature dependent upto the lowest temperature measured (1.8 K) in all the specimens. These findings suggest that the multistep magnetization anomaly in this system may be characterized by a magnetic correlation length.

# 1. Introduction

In the field of geometrically frustrated magnetism, the spin-chain compound, $Ca_3Co_2O_6$ and its derivatives, crystallizing in a $K_4CdCl_6$-derived rhombohedral structure, are currently attracting a lot of attention due to a wide variety of exotic magnetic, magnetodielectric and thermoelectric properties [See, for instance, Refs 1-16 and articles cited therein]. The structure of $Ca_3Co_2O_6$ consists of chains of $CoO_6$ distorted trigonal prism alternating with $CoO_6$ distorted octahedra with the chains separated by Ca ions. The ferromagnetic spin chains (now believed [15] to be modulated along c-axis) are arranged in a triangular fashion in the basal plane with an inter-chain anti-ferromagnetic interaction leading to geometrical frustration. In this compound, as the temperature ($T$) is reduced, long range magnetic ordering takes place at $T_N \sim 24K$. However due to geometrical frustration, the ordering is not complete, giving rise to a 'partially disordered antiferromagnetic' (PDA) state in which two of the three ferromagnetic chains are coupled antiferromagnetically while the third one remains incoherent. On further lowering of $T$, a complex magnetic state involving freezing of spins is observed around 7-9 K.

An important characteristic of this compound that is of current theoretical and experimental interest [3-7; see articles cited in references 9, 10 and 16] is the equally-spaced multiple steps for certain rates of variation of magnetic field ($H$) in the plot of magnetization ($M$) versus $H$ curves at low temperatures (<< 7 K). The exact origin of this puzzling finding is a matter of intense debate in the current literature. It appears that this material is characterized by interesting dielectric [17] and thermoelectric [18] properties with potential applications. It is therefore important to identify a route to synthesize fine particles for any possible applications and to understand the magnetization behavior. The polycrystalline materials employed for investigations in the literature have been synthesized through a solid state route. We therefore considered it worthwhile to prepare this compound in the form of fine particles with dimensions less than micron range through a chemical route and to see how the properties of the particles thus prepared are modified in reduced dimensions.

# 2. Experimental details

The chemical route to synthesize the fine particles of $Ca_3Co_2O_6$ is a 'pyrophoric method' [19], using high purity $CaCo_3$ and $Co(NO_3)_2.6H_2O$. We have employed an aqueous solution of the requisite amounts of these compounds. With the intention of varying the particle size, triethanolamine (TEA) was added to these solutions in such a way that the metal to TEA ratio in the starting solution is maintained at 1:4 (N1), 1:6 (N2), 1:12 (N3) and 1:15 (N4) [20]. The solutions were continuously heated on a hot plate at about 200 C and when complete dehydration occurs, a voluminous, organic based, black precursor powder is left behind. This powder for each ratio was heat-treated at 900 C in air for 4 hrs to get $Ca_3Co_2O_6$ fine particles. The samples were characterized using x-ray powder diffraction (XRD) and we did not find any extra line due to any impurity phase within the detection limit of this technique (see figure 1 for a typical XRD pattern). We do not find any observable variation of the XRD line-widths among the samples N1 to N4, as though the particle sizes are similar in all these cases. On the basis of these XRD patterns, we attempted to get particle sizes from Williamson-Hall plots, but we found that the plots are not linear. This could mean that the particle sizes or shapes are not uniform. In order to get a better insight into the particle size/shape, we have performed field-emission scanning electron microscopic (FE-SEM) studies on all these specimens. Typical SEM images are show in figure 2 and it is clear that the particles in all cases are made up of rods which are fused together in a random manner. A careful look at the SEM pictures reveals that the number of thinner rods increases as one moves from N1 to N4 (thickness: ~500-1000 nm for N1 and <<500 nm for N4). It is thus clear that there is a subtle variation in the thickness of the rods with increasing TEA concentration. We have also carried out energy dispersive x-ray (EDAX) measurements to confirm the compositions. Dc *M(T)* measurements (1.8 to 300K) were carried out using a commercial SQUID magnetometer (Quantum Design) and the same instrument was used to do the *ac* susceptibility ($\chi$) measurements. For *M(H)* (upto 120 kOe) measurements at selected temperatures, a commercial vibrating sample magnetometer (Oxford Instruments) was used; the rate of change of *H* is kept at a value (4 kOe/min) at which the *M(H)* steps in single crystals were reported.

## 3. Results and discussions

Figure 3 shows the *T*-response of zero field cooled susceptibility ($\chi_{zfc}$) and field cooled susceptibility ($\chi_{fc}$) measured in a field of 100 Oe for all the specimens. As known in the literature, the feature due to onset of magnetic ordering around 24 K, a maximum at a lower temperature in $\chi_{zfc}$ and a marked bifurcation of $\chi_{zfc}$ & $\chi_{fc}$ curves at the peak are distinctly noted for all samples. It is to be stressed that in single crystals [3] as well as in nanocrystals synthesized by high-energy ball-milling [9], the peak in $\chi_{zfc}$ appears at about 12 K, whereas in the present specimens, it occurs at a lower temperature (~8 K). This implies that the 8K-shoulder in figure 5 of Ref. 3 noted for single crystals must bear some relevance, possibly relating to a crossover in relaxation mechanism around this temperature (see below). Another point to be noted is that the bifurcation temperature of the $\chi_{zfc}$ and $\chi_{fc}$ curves decreases from 13 K for N1 to 10 K for N4 as though a reduction in the rod diameter has an effect on the bifurcation temperature. In order to check whether there is a change in magnetic moment in Co or in exchange interaction strength, dc $\chi$ measurements in 5 kOe were performed up to 300 K. Inset of figure 3 shows the plot of inverse $\chi$ verse T over a wide *T* (150-300K) range for all samples. The effective moment (found from the Curie-Weiss fitting of the linear region) falls in the range 5.1-5.3 $\mu_B$/formula-unit, which is nearly the same as that of the bulk. The paramagnetic Curie temperature is also the same as in bulk (36-40 K). These results imply that the Co spin and valence as well as exchange interaction strengths are not dramatically influenced for the size range of the particles attained in the present studies.

In figure 4a, we show the *M(H)* isotherms at 1.8 K. It is apparent that the sharp jumps in *M* with 12 kOe interval noted for the single crystalline form [3, 5] for the forward isotherm curve tend to smoothen out in our specimens. The gradual weakening of steps as one moves from N1 to N4 is more clearly visible from figure 4b in which *dM/dH* versus *H* is plotted. This reveals that the thickness of the particle has a decisive role on the steps. The effect is very dramatic for the step appearing near 36 kOe and it is thus interesting to note that there is a subtle difference in the dependence of each of these steps on thickness. The fine particles prepared through the present route have a behavior intermediate between that of single crystals and the nanoparticles synthesized by ball-milling. Clearly, this observation endorses our previous assertion that the multi-step *M(H)* feature for the bulk form is characterized by a length scale possibly decided by magnetic correlation length (MCL). The MCL along *c*-axis has been experimentally determined to be of the order of 550nm whereas it is of the order of 18nm along the basal plane [13-15]. When the particle dimensions fall in this critical range, one would expect that the *M(H)* steps tend to weaken. Though it is not clear whether each of the rods are single crystalline oriented along c-axis, for N4, for which there are more rods with a thickness less than 550 nm, the *M(H)* steps are most broadened. Needless to state that the *M(H)* curve at 15 K reveals a plateau (at 1/3 of saturation moment) followed by a jump at 36 kOe characteristic of PDA structure and this feature remains unchanged for all specimens. Therefore, the modification of *M(H)* seen at 1.8 K is genuinely an effect of a reduction in the particle dimension.

We have also performed ac $\chi$ as well as isothermal remnant magnetization ($M_{IRM}$) measurements on the specimens. For experimental details, the reader may see Ref. 9. The behaviors of $M_{IRM}$ as well as the frequency (*f*) dependence of *ac* $\chi$ in all specimens are essentially the same as that seen in nanocrystals obtained by ball-milling [9] and hence we do not discuss all the data in this article. We present imaginary susceptibility ($\chi''$) versus *f* (Fig. 5a) and $M_{IRM}$ (Fig. 5b) behavior for two specimens only to enable us to focus on the temperature dependence of spin relaxation time ($\tau$). For temperatures above 6 K, we have derived this parameter from the peak-frequency in the plot of imaginary susceptibility ($\chi''$) versus *f*. We observe (see figure 6) thermally activated behavior of $\tau$ above 6 K. For lower temperatures, such a peak in $\chi''(f)$ appears to occur (see figure 5a) at lower frequencies not accessible with our SQUID magnetometer. Therefore, we have derived $\tau$ from $M_{IRM}$ at these temperatures. $M_{IRM}$ curves decays with time (*t*) and a fit to a stretched exponential function of the form, $M_{IRM}$ = a + b exp $[-(t/\tau)^\beta]$ (a, b, and $\beta$ are constants), yields a value of $\tau$ that gradually increases with the decrease in *T* (see figure 6). Such a variation of $\tau$ is most prominent for N4. Thus, the *T*-independence of $\tau$ reported for single crystals [6] below 8 K is not observed in these specimens mimicking the behavior in ball-milled nanospecimens [9]. These results possibly mean that the concept of quantum tunneling may not be a good description to describe the low temperature properties of this compound, supporting the recent conclusion of Soto et al [16].

## 4. Conclusions

We have found a method to synthesize fine particles of a geometrically frustrated spin-chain system, $Ca_3Co_2O_6$, by a pyrophoric method and studied its magnetic behavior. The specimens obtained are in the form of fused rods with an irregular shape of the particles for all concentrations of TEA, but the rod thickness tends to decrease with an increase in TEA concentration. While the magnetic transitions and the magnetization behavior are qualitatively the same as in single crystalline form, the magnetization steps tend to smear out with reducing rod thickness as though there is a length scale involved in the appearance of these steps.

**Acknowledgements**

One of us would like thank N.R. Selvi and G.U. Kulkarni, Jawaharlal Nehru Center for Advanced Scientific Research, Bangalore, India, for SEM measurements. We would like to thank Kartik K Iyer for his help in these studies.

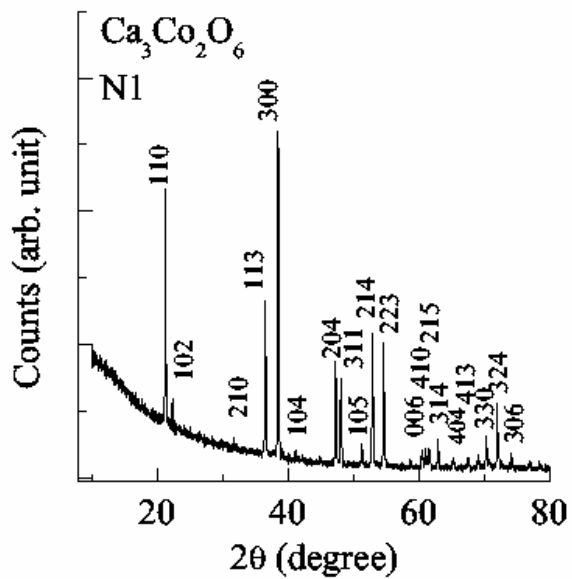

Figure 1: X-ray diffraction pattern (Cu K$_\alpha$) of specimen N1 of Ca$_3$Co$_2$O$_6$.

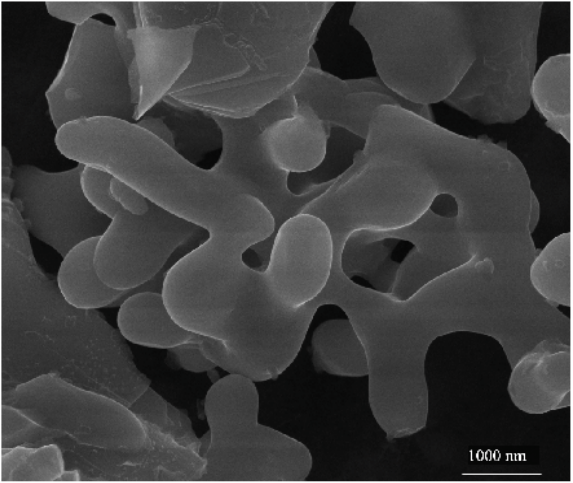
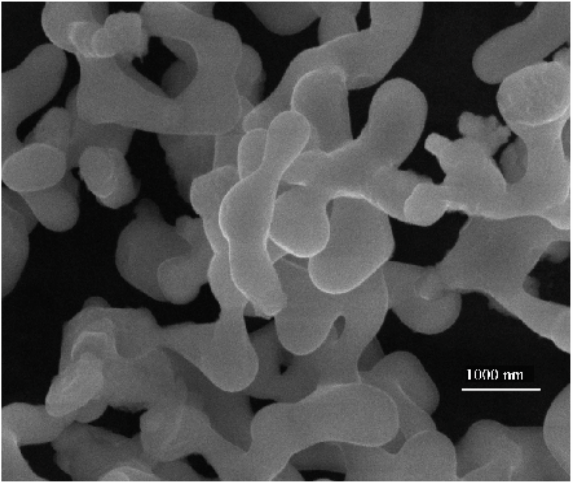

(upper panel) and N4 (lower panel) for $Ca_3Co_2O_6$.

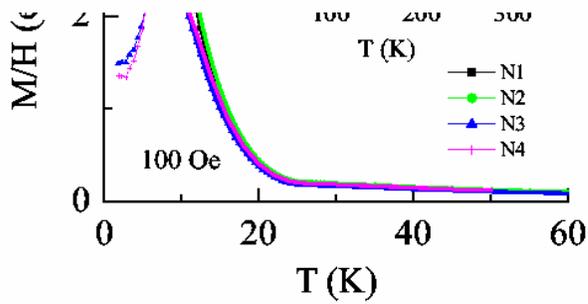

Figure 3:
(color online) Temperature ($T$) dependence (below 60 K) of magnetization divided by magnetic field obtained in a field of 100 Oe for the fine particles N1 to N4 of $Ca_3Co_2O_6$. Data points are shown for zero-field-cooled curves only. Inset: The inverse susceptibility at 5 kOe for all specimens and the dashed straight lines are obtained by Curie Weiss fitting of the data above 150K; the data for N1, N2 and N3 nearly overlap.

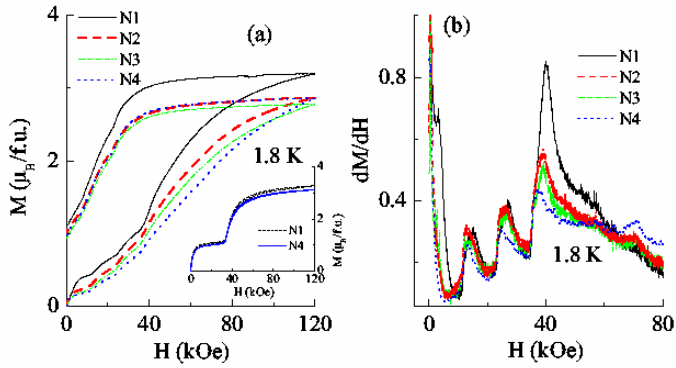

Figure 4:
(color online) (a) *M(H)* isotherms at 1.8 K for all specimens of $Ca_3Co_2O_6$. Inset: *M(H)* isotherms for N1 and N4 samples at 15 K. (b) The field-derivative of *M* as a function of *H* for all the specimens (below 80 kOe).

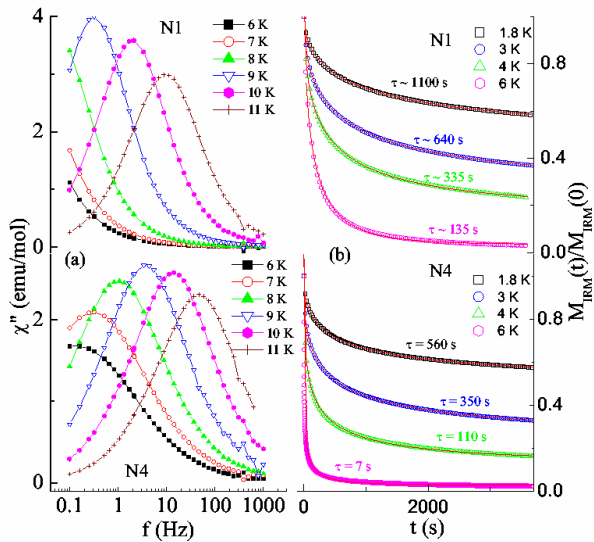

Figure 5:
(color online) (a) Imaginary part of *ac* susceptibility as a function of frequency, and (b) normalized isothermal magnetization as a function of time at selected temperatures for the specimens N1 and N4 of $Ca_3Co_2O_6$.

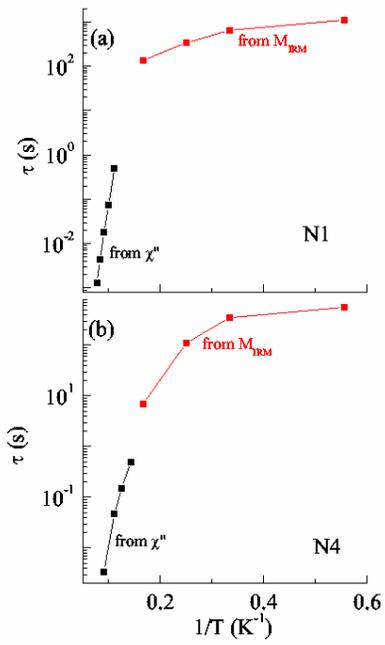

Figure 6:
(color online) (a) Relaxation time ($\tau$) plotted as a function of inverse temperature for N1 and N4 specimens of $Ca_3Co_2O_6$.